# Sulfobetaine-phosphonate block copolymer coated iron oxide nanoparticles for genomic locus targeting and magnetic micromanipulation in the nucleus of living cells

*Fanny Delille,† Elie Balloul,‡ Bassam Hajj,‡ Mohamed Hanafi,¶ Colin Morand,‡,§ Xiang Zhen Xu,† Simon Dumas,†† Antoine Coulon,‡,§ Nicolas Lequeux\*,† and Thomas Pons\*,†*

† Laboratoire Physique et Etude des Matériaux, ESPCI-Paris, PSL Research University, CNRS, Sorbonne Université, UMR 8213, 10, rue Vauquelin, 75005 Paris, France

‡ Laboratoire Physico Chimie Curie, Institut Curie, PSL Research University, Sorbonne Université, CNRS UMR168, 75005 Paris, France

¶ Sciences et Ingénierie de la Matière Molle, UMR 7615, ESPCI Paris PSL─CNRS─Sorbonne Université, 10 Rue Vauquelin, 75005 Paris, France

§ Laboratoire Dynamique du Noyau, Institut Curie, PSL Research University, Sorbonne Université, CNRS UMR3664, 75005 Paris, France.

†† Institut Pierre-Gilles de Gennes, Institut Curie, Sorbonne Université, PSL Research University, 6 rue Jean Calvin, 75005 Paris, France

**Corresponding Authors.** nicolas.lequeux@espci.fr; thomas.pons@espci.fr

ABSTRACT. Exerting forces on biomolecules inside living cells would allow us to probe their dynamic interactions in their native environment. Magnetic iron oxide nanoparticles represent a unique tool capable of pulling on biomolecules with the application of an external magnetic field gradient; however, their use has been restricted to biomolecules accessible from the extracellular medium. Targeting intracellular biomolecules represents an additional challenge due to potential nonspecific interactions with cytoplasmic or nuclear components. We present the synthesis of sulfobetaine-phosphonate block copolymer ligands, which provide magnetic nanoparticles which are stealthy and targetable in living cells. We demonstrate for the first time their efficient targeting in the nucleus and their use for magnetic micromanipulation of a specific genomic locus in living cells. We believe that these stable and furtive magnetic nanoprobes represent a promising tool to manipulate specific biomolecules in living cells and probe the mechanical properties of living matter at the molecular scale.

Magnetic iron oxide nanoparticles (IONPs) have found a broad range of applications in the field of biology, biotechnology and medicine, ranging from magnetic micromanipulation to magnetic resonance imaging and magnetic hyperthermia.[1]

Magnetic micromanipulation is quickly becoming an important tool to understand cellular and intracellular interactions in living cells.[2] Indeed, mechanical properties of biomolecules play a major role in how they organize and respond to external stimuli. The application of an external magnetic field gradient on magnetic micro- and nano-particles enables measuring these mechanical properties and manipulating biomolecules and organelles cells in living cells. For example, IONPs conjugated to specific proteins may be used to create intracellular protein gradients, in order to guide actin polymerization and neurite growth.[3–9] Beyond manipulating the local concentrations of proteins, IONPs may be targeted to bind to specific proteins, located on the outer plasma membrane, such as mechanosensitive receptors, [10–14] which can then be activated to induce cascade biological responses, even though the exact mechanism remains debated.[15] They have also been used to target and manipulate membrane adhesion proteins such as E-cadherins, in order to guide receptor concentration at specific locations on the membrane and interrogate their interactions with the submembrane actin cytoskeleton. [16,17]

Application of forces to micro-manipulate specific proteins using IONPs has been restricted so far to extracellular applications. Intracellular targeting of IONPs represent an additional challenge, due to the crowded and complex environment and the potential nonspecific interactions with intracellular biomolecules and components. Indeed, these may induce NP aggregation, which would restrict their diffusion if their size becomes larger than the size of the cytoplasmic mesh (50-75 nm).[18] It would also cause IONPs to stick to cytoplasmic components and prevent the desired biomolecular recognition.[18,19] A protein-derived magnetic particle has been recently developed, based on an iron oxide particle grown inside a ferritin cage that was fused with additional proteins and further conjugated with an antifouling poly(ethylene glycol) layer. This tool was recently demonstrated to manipulate specific intracellular organelles[20] and genomic loci in the nucleus of living cells.[21]

Being able to target IONPs obtained by thermal decomposition synthesis to specific intracellular biomolecules would greatly broaden the range of application of micromanipulation techniques, since these are much more versatile than IONPs grown inside ferritin cages. Indeed, IONPs may be easily synthesized with high monodispersity in a broad range of sizes, enabling fine control of the magnetic properties, reaching higher magnetization and thus applied magnetic force. The functionalization of IONPs is also in principle more compact and more versatile than a protein cage, with the ability to conjugate different proteins and vary the nature of the targeted biomolecules easily, add fluorescent markers or other small reporters, etc. However so far this application has been limited due to the lack of efficient surface hydrophilic ligands, which should (i) strongly bind to the surface atoms of IONPs and not desorb, even in the complex intracellular environment, (ii) avoid any nonspecific interaction with intracellular biomolecules to freely diffuse and (iii) enable bioconjugation and targeting to specific biomolecular targets.

Sulfobetaine polymers have been demonstrated to enable protein repulsion and stabilization of gold nanoparticles and quantum dots in physiological media.[19,22,23] We have previously demonstrated that block copolymer ligands composed of a multidentate anchoring block and a poly(sulfobetaine) hydrophilic block also enabled unhindered diffusion of semiconductor quantum dots in the cell cytoplasm and their efficient biomolecular targeting.[19,24] This result has not been however generalized to other types of nanoparticles, and was never extended to targeting in the nucleus of living cells. Here, we report the design and synthesis of a block copolymer ligand adapted to IONPs, composed of poly(phosphonic acid) and poly(sulfobetaine). We demonstrate the excellent stability of this polymer on the IONP surface, even under binding competition conditions, and the absence of any significant nonspecific interactions of these magnetic nanoparticles in model biofluids *in vitro* or in the cell cytoplasm. Finally, we develop a functionalization scheme to target these IONPs to a specific genomic locus on chromosome 1 in a human osteosarcoma cell line and demonstrate the ability to apply mechanical forces and micro-manipulate this chromosome in the nucleus of living cells.

**Results and discussion:**

Iron oxide nanoparticles (IONPs) were synthetized by thermal decomposition of iron oleate in octadecene, according to the protocol developed by Park et al.[25] The amount of excess oleic acid was chosen to yield nanoparticles with a diameter of either 7.9 ± 0.9 nm or 10.3 ± 1.7 nm, as measured by transmission electron microscopy (Supporting Information, SI, Figure S1), and thereafter referred to as 8 nm or 10 nm IONP for simplicity. The final nanoparticles are capped with oleate ligands and dispersed in hexane. The saturation magnetization was measured at 71 emu/g at room temperature for the 10 nm IONP (SI Figure S2), close to the value expected for maghemite (78 emu/g).

Following a strategy previously developed to functionalize semiconductor quantum dots,[19,24] we designed a block copolymer ligand composed of two blocks (Figure 1A). The first block is a hydrophilic copolymer composed of sulfobetaine monomers (SPE) to ensure colloidal stability and limit nonspecific interactions with biomolecules, mixed statistically with azide ($N_3$)-terminated monomers to enable bioconjucation using bio-orthogonal azide-alkyne click chemistry.[26] The second block is composed of multiple anchoring functions, in order to enhance its stability on the nanoparticle surface thanks to the multidentate effect. As an anchoring function, phosphonic acids (PA) were selected for their good affinity for iron surface atoms in iron oxide nanoparticles, as previously reported.[27–31] The first block, p(SPE-N$_3$), was synthesized by RAFT polymerization as a statistical copolymer, from a mixture of a sulfobetaine-containing methacrylate monomers (SPE) and N$_3$-terminated methacrylamide monomers, in the presence of the RAFT agent 4-cyanopentanoic acid dithiobenzoate (CADB) and azobisisobutyronitrile (AIBN) as an initiator in an acetic acid/water mixture (SI Figure S3). This first block was purified by precipitation and the second block was polymerized in water using phosphonate-terminated methacrylate and 2,20-azobis(2-amidinopropane) hydrochloride (V50) in water. The resulting block copolymer, p((SPE-N3)-*b*-PA), was purified by precipitation and dried. The molecular weights of the different blocks were estimated from GPC and $^1$H NMR, and may be adjusted by varying the monomer and CADB molar ratios.[19,24] The results presented hereafter are obtained with a polymer composed of a first block containing 33 SPE and 5 N$_3$ monomers, on average, for a total molecular weight of 9,300 g·mol$^{-1}$, as estimated from GPC and NMR, with an index of polydispersity of 1.06. The anchoring block is composed of ca. 5 PA monomers, as estimated by $^1$H NMR and GPC (SI Figures S4-S6).

Since sulfobetaine polymers and oleate-capped IONPs are not dispersible in the same solvents, we performed the ligand exchange through a two-step procedure (Figure 1B), starting with ligand stripping using NOBF$_4$.[32] The p((SPE-N$_3$)-*b*-PA) copolymer is then added in water. The pH of the mixture is acidic (pH ≈ 3.5), due to the partial deprotonation of PAs (pKa$_1$ ≈ 3). At this pH, the IONP surface is positively charged (pI ≈ 6.8),[33] which favors the binding of the negatively charged, partially deprotonated, PAs.[29] The pH is then increased to pH ≈ 7-7.5, in order to completely deprotonate the PA (pKa$_2$ ≈ 5-6) and improve its binding strength on the IONP surface. The resulting IONPs functionalized with the copolymer, denoted (SPE-N$_3$)-IONP, are then purified by ultracentrifugation in order to eliminate potential aggregates and excess polymer and recover individual functionalized IONPs. The absence of aggregates is confirmed by dynamic light scattering, with measured hydrodynamic diameters of ca. 20-22 nm for 8 nm diameter IONPs and ca. 23-25 nm for 10 nm IONP (Figure 1C). We then tested the stability of (SPE-N$_3$)-IONPs at different pHs between 5 and 9, and in 5 M NaCl. Even after one month of incubation in these conditions, the nanoparticles remained colloidally stable, as shown by the absence of formation of any pellet under high-speed centrifugation (Figure 1D, S7). This confirms that the polysulfobetaine coating imparts an excellent colloidal stability in these conditions, consistently with what has been previously reported on other nanoparticles.[24,34,19]

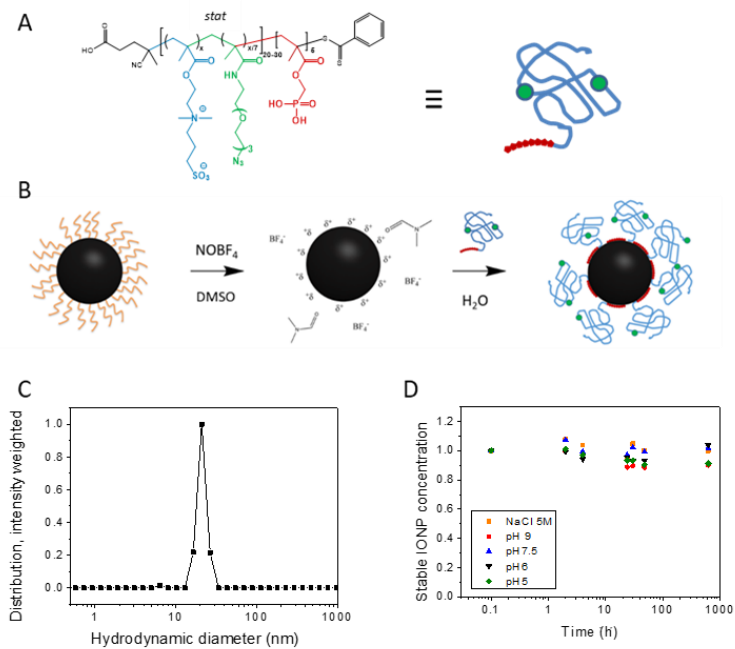

Figure 1: A. Schematics of the copolymer ligand with a phosphonic acid anchoring block (red), and a hydrophilic statistical block composed of sulfobetaine (blue) and azide (green) monomers. B. Ligand exchange procedure; C. Hydrodynamic size distribution obtained by Dynamic Light Scattering for 8 nm diameter IONP; D. Colloidal stability of (SPE-N$_3$)-IONPs in HEPES or MES buffers at different pH and in 5 M NaCl solution: the IONP solutions were centrifugated at high

speeds to eliminate any aggregates, and the concentration of remaining stable particles in the supernatant was measured by absorption spectroscopy.

In order to quantify the anchoring stability of the PA block on the IONP surface, we ran a competition experiment (Figure 2A). First, azide groups within the copolymeric ligands were reacted with dibenzoyl cyclooctyne (DBCO)-fluorophores after functionalization of the IONPs. After removing the excess unbound fluorophores, the absorption spectrum presents the signature of the IONPs and the fluorophores (Figure 2B), which enables an estimation of the average ratio of fluorophore per IONP, with an initial value of 12. Then, a 500x fold excess of unlabeled p(SPE-b-PA) ligands was added, such that if any of the initial fluorophore-labeled ligands desorbed from the surface over time, they would be replaced by an unlabeled one. Measuring the absorption spectrum of solution samples over time after ultrafiltration, to separate solution-phase polymers from IONPs, thus enabled evaluating the desorption kinetics of the polymer ligands from the surface. The fraction of desorbed ligands was estimated at less than 3% after one week, at room temperature ($k_{off} < 10^{-7}$ s$^{-1}$, Figure 2C). This confirms the high stability of p(PA) anchoring block, a crucial feature for the long-term stability of functionalized IONPs in complex media such as the cell cytoplasm or nucleus.

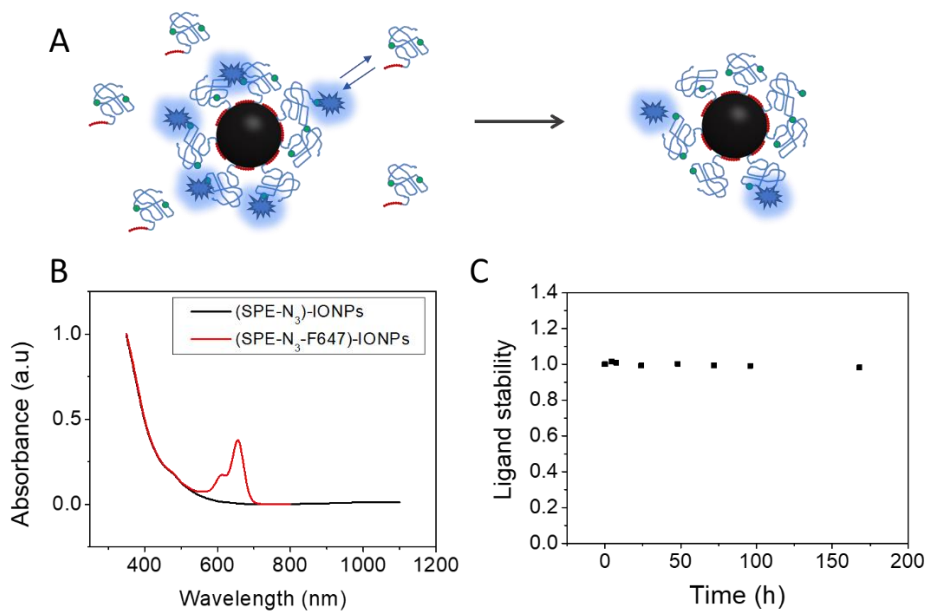

Figure 2. A. Schematics of the ligand stability assay. B: Normalized absorption spectrum of IONP before and after functionalization with FP647 fluorophore. C: Normalized amount of labeled ligand remaining on the IONP surface as a function of the time of incubation with a large excess of unlabeled ligands.

We then investigated the antifouling properties of these fluorophore-labeled IONPs in the presence of an increasing concentration of bovine serum albumin (BSA, hydrodynamic diameter ≈ 7 nm) as a model protein, using fluorescence correlation spectroscopy.[19] Consistently with previous observation on QDs coated with poly-sulfobetaine ligands, the hydrodynamic size did not show any significant increase even in millimolar concentrations of BSA, corresponding to the concentration of albumin in blood (Figure 3A). This suggests an absence of any nonspecific interactions between the sulfobetaine-coated IONPs and BSA. In order to test a more complex biological fluid, we then tested mouse plasma, which is composed of hundreds of different proteins with diverse physico-chemical properties, and thus constitutes a more stringent test of nonspecific interactions. Again, no significant changes in hydrodynamic size were observed, suggesting that even in the presence of a broad variety of proteins, the nonspecific interactions are weak or inexistent. If any interaction exists, it does not lead to any IONP aggregation or to a detectable protein corona (Figure 3B).

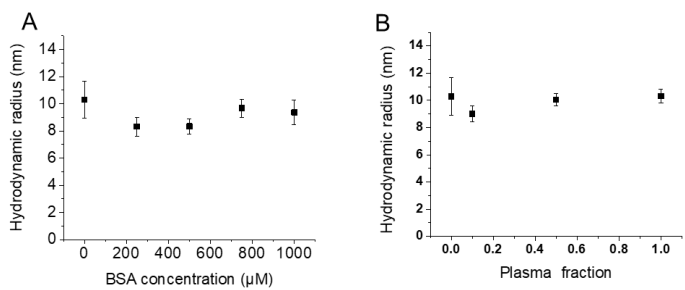

Figure 3. Hydrodynamic radius of (SPE-N3-FP647)-IONPs (inorganic diameter 8 nm) obtained by FCS in HEPES buffered saline (HBS) as a function of BSA concentration (A), and as function of mouse plasma volume fraction (B).

Having demonstrated the excellent antifouling properties of the p((SPE-N3)-PA)-coated IONPs in model biofluids *in vitro*, we then turned to the intracellular medium. Fluorophore-functionalized IONPs were micro-injected in the cytoplasm of RPE-1 cells. Fluorescence microscopy shows a homogeneous repartition of the IONPs in the cytoplasm, without any visible accumulation or agglomeration (SI Figure S8). This suggests that the small size of our IONPs enables a diffusion which is not affected by obstructions from fixed solid structures in the cytoplasm (pore size ≈ 50-75 nm).[18] This is confirmed by FCS (SI Figure S8), with diffusion coefficients in the cytoplasm on the order of a few $\mu m^2 \cdot s^{-1}$. This is consistent with reported values for particles of similar sizes that are freely diffusing without encountering any nonspecific interactions.[18,19,35,36] In order to test the ability of these nanoparticles to be manipulated by a magnetic field gradient inside the cells, we approached a permalloy-coated tip magnetized by an external magnet in the extracellular medium. The IONPs moved rapidly within the magnetic field gradient and accumulated in the area closest to the magnetic tip (SI Movie 1). Even under these conditions, they did not agglomerate and redispersed readily in the cytoplasm when the magnetic tip was removed.

Since these magnetic nanoparticles appear to freely diffuse inside cells without being hindered by nonspecific interactions with biomolecules, we then tested their application for magnetic micromanipulation of genomic loci. Indeed, magnetic micromanipulation at the level of individual loci would be a powerful tool to disturb the complex genome architecture and therefore study the causal links between organization and function of the genome.[37] In these experiments, 10 nm IONPs were selected to obtain higher magnetic force. In order to target IONPs to a specific genomic locus, we used a human U-2 OS cell line modified to present a repetitive array of *tetO* DNA sequences at a single genomic location on chromosome 1[38] and expressing the TetR protein fused with the fluorescent mCherry protein and an anti-green fluorescent protein (GFP) nanobody,[21] such that the TetR-mCherry-antiGFPnb can bind on one side to the *tetO* site and to GFP on the other side. We prepared a DBCO-modified GFP to conjugate to (SPE-$N_3$)-IONPs. The final GFP:IONP ratio was set to 3, in order to minimize both the fraction of unconjugated IONPs and the possibility of multiple GFP binding from single IONP. The IONPs were then functionalized with a FP647 fluorophore to enable their localization by fluorescence microscopy. DLS measurements showed that the GFP conjugation increased the hydrodynamic radius of the IONPs from ca. 23-25 nm to ca. 29 nm. GFP-FP647-IONPs were microinjected into the nucleus of modified U-2 OS cells and imaged by fluorescence microscopy. As shown in Figures 4A and S9, the mCherry channel reveals the location of the TetR protein, with an accumulation in one spot inside the nucleus, corresponding to the *tetO* locus on chromosome 1. The channel corresponding to the FP647 fluorophore, Figure 4B, shows the location of the functionalized IONPs. Interestingly, not all the cells actually express the TetR protein construct, as evidenced by the lack of fluorescence in the mCherry channel (cell #1 in Figure 4). In these cells, the IONPs diffuse homogeneously inside the nucleus, except into nucleoli, without any visible agglomeration, which demonstrates the absence of nonspecific binding. In contrast, in cells expressing the TetR construct (cells #2-6 in Figure 4), the IONPs accumulate onto the *tetO* locus, as evidenced by a colocalization with the TetR localization (Figure 4C, SI Figures S10-S11). The fluorescence intensity ratio between the locus and the rest of the nucleus is typically 50 to 100, demonstrating a strong targeting efficiency. This accumulation occurs within minutes after the nanoparticle microinjection inside the nucleus, as shown by the rapid increase of FP647 fluorescence at this locus (Figure 4D). This demonstrates the rapid and efficient targeting of our GFP-functionalized IONPs, despite the very complex and crowded environment of the living cell nucleus. Comparison between the fluorescence intensity of the locus spot in the FP647 channel with the intensity from individual nanoparticles enables an estimation of the number of IONP on the locus, which varies from cell to cell and ranges from 300 to 3000 (average value of 1400 ± 1000, measured on 33 cells). This variation may be due in part to the different IONP solution volumes that are microinjected in each cell.

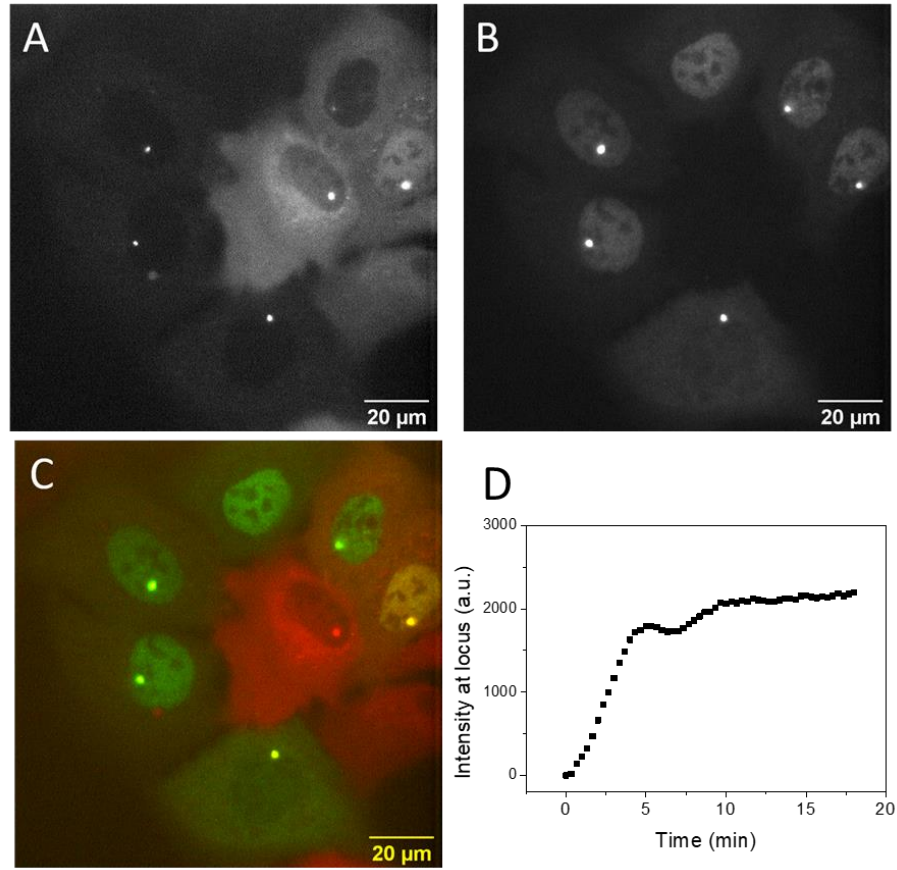

Figure 4. Fluorescence microscopy images of U2-OS modified cells microinjected with 10 nm GFP-FP647-IONP: A. mCherry signal from the TetR-mCherry-GFPnb; B. FP647 from the functionalized IONPs; C. Colocalization of the mCherry signal (red) and the FP647 signal (green); D. Evolution of the FP647 fluorescence intensity at the *tetO* locus with time after microinjection of the functionalized IONPs.

In the absence of external magnetic field, the locus slowly moves, with typical displacements much smaller than 1 μm on the time scale of 10 min, due in part to cell and nucleus movement (SI Figure S12). In order to apply a force on the chromosome, a permalloy coated magnetic tip was magnetized by an external permanent magnet and approached, using a micromanipulator, close to the cell nucleus without touching the cell. As shown in Figures 5A-B (kymograph shown in Figure S13), this induces an immediate displacement of the genomic locus towards the magnetic tip, where the magnetic field is stronger, at speeds in the µm/min range, which in this example ends when the locus reaches the nuclear periphery, without being able to cross the nuclear membrane under these conditions (Supplementary Movie 2). Modeling of the magnetic field around the tip suggests that the forces applied on the locus are in the 1-10 pN range (Figure S14). Magnetic micromanipulation is quite versatile in its application and also enables modulating the applied force over time. When applying a force and then releasing it by pulling away the magnetic tip, different behaviors were observed in different cells. This includes a locus which first moves in response to the force, then remains in position after force release (Figure 5C), or, in contrast, a locus which moves in response to the force and then recoils back when the magnetic tip is removed (Figure 5D and Supplementary Movie 3), suggesting the existence of elastic restoring forces. This cell-to-cell variability in the response of a chromosome to a force calls for further investigation.

Compared to previously described ferritin nanoparticles used in micromanipulation, our functionalized IONPs present several advantages. First, the size of the iron oxide core in ferritin NPs is constrained by the size of the ferritin cage (approximately 8 nm), which limits the magnetization to $4 - 9 \, 10^{-17}$ emu/NP. This limitation is lifted by the use of IONPs, which size can be freely tuned during synthesis. In comparison, our 10 nm IONP present a magnetization of $47 \, 10^{-17}$ emu/NP and enable therefore application of higher magnetic forces, while maintaining an equivalent total hydrodynamic size (29 nm for IONP-GFP vs. 28 nm for ferritin NPs). Second, by construction, the ferritin NPs contain 24 GFP, which may induce crosslinking with several targets, whereas IONP may be functionalized with a small controlled number of targeting proteins. Finally, IONPs coated with this sulfobetaine copolymer are more versatile and can be easily conjugated to other targeting moieties such as biotin, streptavidin…, while retaining their colloidal stability in cytoplasm and nucleus, as demonstrated in Figures S15-S16.

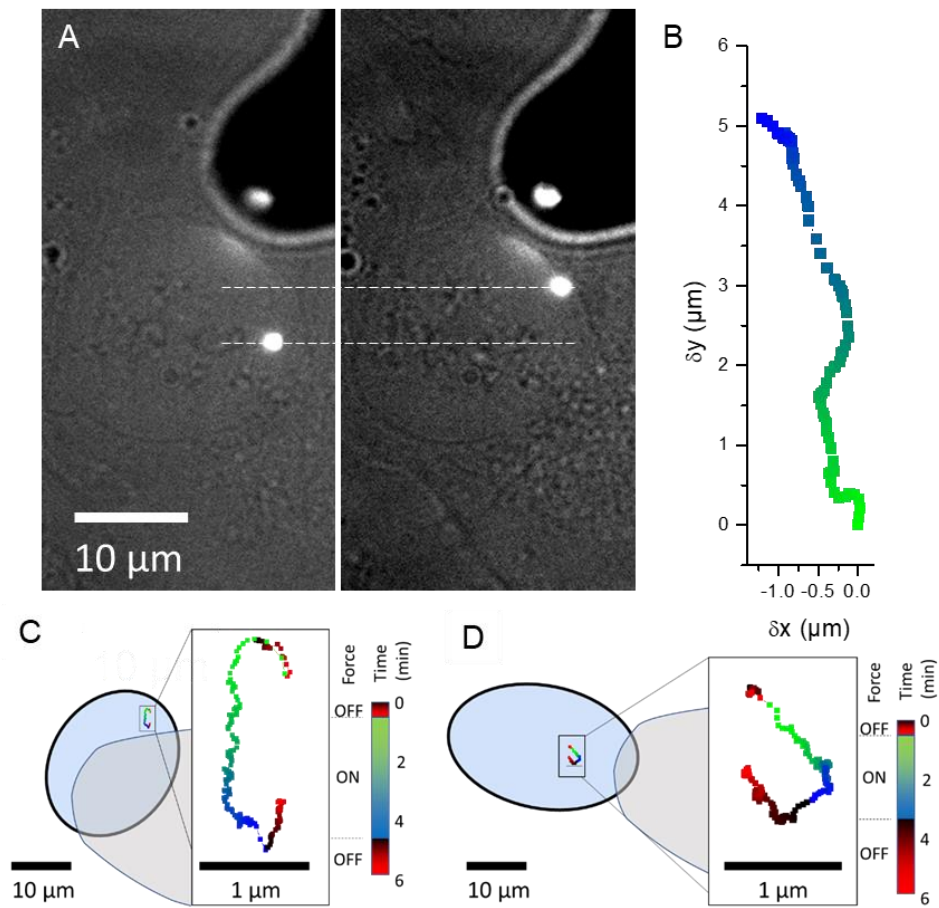

Figure 5. A: Fluorescence and transmission microscopy images showing the position of the locus at t=0 and t=10 min; B: locus displacement: dots represent the displacement of the locus measured every 10 s, color-coded in time from green to blue, total time: 12.16 min. C, D: Locus displacement within the nucleus (shaded in blue) during magnetic micromanipulation in two different cells. Insets show a zoom around the locus. The gray shaded shapes represent the location of the magnetic tip above the cell. The dots in the trajectories are equally spaced in time (2 s) and color-coded during the trajectories (red before bringing the magnetic tip above the cell, gradient from green to blue during application of the magnetic force and from black to red after removal of the magnetic tip).

In conclusion, we have synthesized magnetic iron oxide nanoparticles by a thermal decomposition approach and designed a novel block copolymer ligand for their functionalization, with a block of phosphonic acid for anchoring onto the IONP surface, and a block of sulfobetaine monomers to provide colloidal stability in water and physiological media. We have shown the long-term stability of the polymeric ligand on the IONP surface, as well as the absence of nonspecific interactions with model proteins in model solutions, blood serum or in the cell cytoplasm. We have functionalized these magnetic nanoparticles to target them to a specific genomic locus after intranuclear microinjection. We have demonstrated magnetic micromanipulation of a chromosome in live cells using IONPs. The versatility of the functionalization of these IONPs and their ability to navigate freely in the cytosol suggests that they could be a useful tool for micro-manipulating various biomolecules and biological structures in living cells.

**Acknowledgements.** The authors acknowledge the essential contribution of Maxime Dahan to the design and implementation of this project. The authors thank David Hrabovsky, of the MPBT platform (Physical Measurements at Low Temperatures) of Sorbonne Université, Ilham Aboulfath-Ladid, Julia Ronsch and Maud Bongaerts for magnetization measurements, Brigitte Leridon for helpful discussion and Laure Cordier for ICP-AES measurements.

ASSOCIATED CONTENT

**Supporting Information**. The following files are available free of charge.

Experimental details, Transmission electron microscopy, magnetization measurements, 1H NMR and gel permeation chromatography of the copolymer ligands, cytoplasmic fluorescence correlation spectroscopy, motion of the genomic locus in absence of applied magnetic force (PDF);

Supplementary Movie 1: Micromanipulation of IONPs in the cytoplasm.

Supplementary Movie 2: locus displacement from Figure 5 (AVI);

Supplementary Movie 3: locus displacement from Figure 5 (AVI).

Supplementary Movie 4: Free diffusion of biotin-FP647-IONP in the nucleus (AVI). Frame interval 200 ms, exposure 100 ms, play speed 25x.

Supplementary Movie 5: Free diffusion of biotin-FP647-IONP in the cytoplasm (AVI). Frame interval 200 ms, exposure 100 ms, play speed 25x.

Supplementary Movie 6: Single IONPs diffusing in the cytoplasm (AVI). Frame interval 155 ms, exposure 100 ms.

AUTHOR INFORMATION

**Corresponding Author**

Nicolas Lequeux - Laboratoire Physique et Etude des Matériaux, ESPCI-Paris, PSL Research University, CNRS, Sorbonne University, UMR 8213, 10, rue Vauquelin, Paris, France

nicolas.lequeux@espci.fr

Thomas Pons - Laboratoire Physique et Etude des Matériaux, ESPCI-Paris, PSL Research University, CNRS, Sorbonne University, UMR 8213, 10, rue Vauquelin, Paris, France

thomas.pons@espci.fr

**Author Contributions**

The manuscript was written through contributions of all authors. All authors have given approval to the final version of the manuscript.

**Funding Sources**

The authors acknowledge funding from the Agence Nationale de la Recherche (France, ChroMag project ANR-18-CE12-0023-01)

**Notes**

The authors declare no competing financial interest.

TOC Graphic.

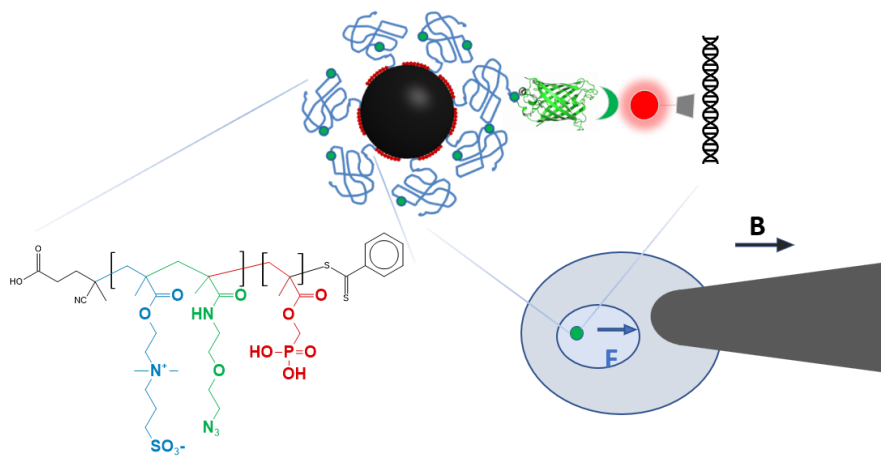